\documentclass[twocolumn]{revtex4}
\usepackage{amssymb}

\input{tcilatex}

\begin{document}

\title{The attainable superconducting $T_{c}$ in a model of \\
phase coherence by percolation.}
\author{D.Mihailovic$^{1}$, V.V.Kabanov$^{1}$ and K.A.M\"{u}ller$^{2}$ \\
$^{1}$Jozef Stefan Institute, Jamova 39, 1000 Ljubljana, Slovenia\\
$^{2}$Physik-Institut der Universit\"{a}t Z\"{u}rich, 8057-Zurich,\\
Switzerland}

\begin{abstract}
The onset of macroscopic phase coherence in superconducting cuprates is
considered to be determined by random percolation between mesoscopic
Jahn-Teller pairs, stripes or clusters. The model is found to predict the
onset of superconductivity near 6\% doping, maximum $T_{c}$ near 15\% doping
and $T_{c}\simeq T^{\ast }$ at optimum doping, and accounts for the
destruction of superconductivity by Zn doping near 7\%. The model also
predicts a relation between the pairing (pseudogap) energy and $T_{c}$ in
terms of experimentally measurable quantities.
\end{abstract}

\maketitle

\newpage The occurrence of pairing and a phase coherent superconducting
state in the cuprates is thought to occur independently, especially in the
so-called underdoped state. This is contrary to the behavior in BCS
superconductors, where they occur simultaneously. This enables us to discuss
the mechanism for the establishment of phase coherence in itself, starting
with pre-formed real-space pairs. In this letter we investigate the
possibility of the formation of a macroscopic superconducting phase-coherent
state by the spreading of phase coherence by percolation (PCP), calculating
the attainable $T_{c}$ and some of the critical points in the phase diagram.

We assume that superconductivity in the cuprates arises when pre-formed
pairs acquire mutual phase coherence. The pairs are considered to be
described by the mesoscopic Jahn-Teller (MJT) pairing model, which gives a
detailed mechanism for pair formation arising from degeneracy of hole
orbitals\cite{MV,MJT}. In this model the size of each pair $l_{p}$
(including deformation of the surrounding lattice and spins) is less than or
equal to the coherence length $\xi _{s}$ but larger than the lattice
constant $a,$ i.e. $a<l_{p}\lesssim \xi _{s}.$ The percolation model
proposed here is considered primarily for the underdoped state which is
mixture of Bosonic pairs and Fermionic excitations\cite{AM,AKM}, separated
by a pairing energy gap $E_{p}$\cite{Gantmakher}.

Let us now assume that macroscopic phase coherence is established when the
percolation threshold is reached and pairs touch each other to form a phase
coherent path throughout the crystal. The situation considered is \emph{%
dynamic}\cite{review}, where the pair presence fluctuates statistically on a
timescale given approximately by $\tau \simeq \hbar /E_{p}\exp (E_{p}/kT).$
Figure 1a) shows a \emph{snapshot}{\huge \ }of the CuO$_{2}$ planes filled
by MJT pairs for 6\% doping at $T=0$ (all pairs in the ground state). The
average size of the pairs is given by the experimentally determined value of 
$\gamma =\pi /l_{p},$ where $\gamma $ is the range of the interaction in $k$%
-space as given by the inelastic neutron scattering (INS) experiments\cite
{Egami}. The area occupied by each pair in the CuO$_{2}$ plane is: 
\begin{equation}
V_{p}\simeq \pi l_{p}^{2}=\pi \left( \frac{\pi }{\gamma }\right) ^{2}
\end{equation}
The color (density) depicts the amplitude of the lattice deformation caused
by the paired carriers predicted by the MJT model. The situation depicted in
Figure 1a) can be described well by the 2D bond percolation (BP) model shown
in Figure 1b), whereby each possible site for a pair is depicted by a dot,
and the occupancy of the site is depicted by a bond to the next dot.

The percolation threshold for the BP model in 2D is known to occur when the
''volume fraction'' $F_{V}=1/2$\cite{BP}. The ''volume fraction'' occupied
by the pairs at temperature $T$ and doping density $x$ is then given by:

\begin{equation}
F_{V}=\frac{1}{2}xn_{p}(T)V_{p}/V
\end{equation}

where $n_{p}(T)$ is the thermal pair occupancy at temperature $T$ and $%
V=a^{2}$ is the volume of the unit cell, and $a$ is the in-plane lattice
constant\cite{tetragonal}. The factor 1/2 arises because two doped carriers
are required to create one pair.

Let us calculate the critical pair density at the percolation threshold at $%
T=0$, which corresponds to the onset of superconductivity on the underdoped
side (point \textbf{A} in Fig.2). In this case all the carriers are paired
in the ground state and $n_{p}(T=0)=1,$ so $x_{c}=2F_{V}V/V_{p}$. Assuming $%
F_{V}=1/2$ and that $\gamma \simeq \pi /2a$ from the INS experiments\cite
{Egami}, we obtain $x_{c}=0.08.$ The experimental value for the onset of
superconductivity for most cuprates is indeed very close, being $0.06\pm
0.01 $ for La$_{2-x}$Sr$_{x}$CuO$_{4}$ \cite{xc} and $x_{c}=0.05\pm 0.01$
for YBCO \cite{xc2} for example. (Note that the insulator-to-metal
transition does not necessarily coincide with this point, since the
percolation threshold for metallic behaviour may be different\cite{Mesot}).

Next, let us consider the situation near optimum doping. Here, the optimum
carrier concentration is typically $15\%.$ From (2), at $T=0$ and $%
x_{opt}=0.15$ (point \textbf{B} in Fig 2) we obtain $F_{V}=0.94.$ Thus near
optimum doping, the CuO$_{2}$ planes are fully occupied by pairs. Further
doping results in overlapping pairs\cite{AM,Mott} and the formation of more
spatially extended metallic regions\cite{MV}.

However, for $T>0$, the population of pairs is reduced, and at some finite
critical temperature $T_{c}$, the percolation threshold is reached (point 
\textbf{C} in Fig. 2). Using Eq. 2 \ with $F_{V}=1/2$ and the experimental
value of $\gamma ,$ we obtain a pair occupancy $n_{p}(T)\simeq 0.53.$ This
means that near optimum doping, at $T_{c}$ the pair occupancy is
approximately half, which is also the point where $kT^{\ast }\simeq E_{P}$
(We follow the convention that the temperature at which the thermal energy
becomes comparable to the pair binding energy is called the pseudogap
temperature $T^{\ast }$). Thus the percolation model predicts that $%
T_{c}\sim T^{\ast }$ near optimum doping, which is indeed the case in most
cuprates\cite{Demsar}.

Considering the population of pairs as a function of temperature in the MJT
model\cite{MJT,excitations}, we can calculate $T_{c}$ as a function of
doping. We start with the expression for the number of pairs $n_{p}(T)\;$at
temperature $T$ predicted by the MJT model\cite{MJT}: 
\begin{equation}
n_{p}(T)\simeq 1-K(E_{p}/kT)^{-1/2}\exp {(}-E_{{p}}/kT{).}
\end{equation}

$E_{p}$ is the pair binding energy, and $K$ is a constant describing the
density of unpaired states. Substituting into (2) and solving for $T$, the
temperature $T_{c}$ where $F_{{V}}{=1/2}$ is approximately given by: 
\begin{equation}
T_{c}(x)/E_{p}\simeq 1/\ln (K\frac{x}{x-x_{c}}).
\end{equation}

To test this relation, in Figure 3a) we plot the predicted $T_{c}$ using
experimental values of $E_{p}$ from quasiparticle recombination data\cite
{Demsar} on YBCO as a function of doping and using the value of $K=200$
obtained from experimental fits of the $T$-dependence of Raman and neutron
scattering data\cite{MJT}. Remarkably the magnitude of $T_{c}$ is near 90K
near optimum doping $x=0.15$. A universal ''pseudogap'' ratio of $E_{p}\,$/$%
T_{c}$ as a function of$\ $doping $x$ given by Eq. (4) is plotted in Figure
3b). Over most of the doping range it is near 6 and increases near the
percolation threshold as $x\rightarrow x_{c}.$ Here the value of $E_{p}\,$/$%
T_{c}$ depends only on $K$ and $x_{c}$, which can be determined
independently from many different experiments\cite{MJT,excitations,Demsar}
and is in agreement with measured values from single-particle tunneling
experiments\cite{Deutscher}.

In the model above we have not discussed the mechanism for establishing
phase coherence between pairs (or clusters of pairs). Since typically the
pair size is approximately equal to the coherence length $l_{p}\simeq \pi
/2\gamma \simeq \xi _{s},$ we describe the mechanism in terms of pair
Josephson tunneling, as was already discussed briefly in ref. \cite{MV}. The
pairs together with their deformation are considered immobile on the
timescale of $\hbar /kT_{c}$, which makes percolation relevant\cite{BEC}.
However, the charge carriers can tunnel from one pair region to the next in
pairs, forming a phase coherent state throughout the crystal. This tunneling
occurs resonantly between the Bosonic states of each MJT pair-region,
without carrying the deformation itself. The tunneling amplitude is
independent of the presence of other particles within the deformed
pair-region, because the deformed regions are sufficiently large to
accommodate more than 2 particles. However, if the final cluster is
populated by a pair, the Coulomb charging energy is given by a Hubbard term $%
U\sim 4e^{2}/(\epsilon _{0}l_{p})$ and we may describe the system in terms
of a Boson-Hubbard model imposed on a pseudo-random bond-percolation
network. (This is somewhat different from the scenario for superconductivity
previously considered in the Boson-Hubbard model on a regular lattice\cite
{BH}.)

The simple percolation model ignores completely the effects of thermal phase
fluctuations, which are known to be important in cuprates and other systems
such as charge-density wave systems\cite{Gruner}. The effect of fluctuations
is to reduce the $T_{c}$, especially in the strongly underdoped region,
where the superfluid density and the phase stiffness are small\cite{EK}.
This could explain the discrepancy between the predicted $T_{c}$ and the
actual $T_{c}$ for small doping (Fig.3a).

We can apply the same percolation arguments to the discussion of the role of
metal impurities such as Ni and Zn which replace Cu in the CuO planes. By a
modified ''swiss-cheese'' type of argument\cite{Nachumi}, if we ignore
impurity aggregation, the real-space area occupied by each impurity is of
the order of $\pi \left( \frac{\pi }{\gamma }\right) ^{2}.$ This space is
unavailable for the formation of pairs. (The justification for this comes
from many different experiments, the most direct being the scanning
tunneling microscope measurements of Pan et al (2000)\cite{Pan} and muon
spin relaxation\cite{Nachumi}.) By equation (2) the critical concentration
of impurities for the total destruction of superconductivity (i.e.
percolation threshold at $T=0)$ is $x_{c}^{Zn}=8$\%, close to the 7\%
typically found experimentally for Zn\cite{Nachumi}. Of course different
ions may inhibit the formation of mesoscopic pairs over slightly different
distances and the exact value of the critical impurity concentration will
vary.

The effect of increasing carrier density is to make aggregation of pairs
into stripes more and more probable\cite{MV,MJT}. Even at the percolation
threshold near 6\% doping (Fig. 1a), the planes are substantially occupied
and it is natural that the formation of longer stripes will occur,
especially at low temperatures. The additional energy gain associated with
the formation of such stripes will be mainly associated with the sharing of
the deformation and the energy scale for the aggregation is expected to be
significantly smaller than the pairing energy (pseudogap). Consequently the
formation of stripes (such as is shown in Fig.1a)\ will occur at
temperatures below $T^{\ast }$, and their density will be determined by
statistics. However, if these larger stripes or clusters are metallic, they
occupy space, but \emph{do not} contribute to the superconducting volume
fraction. Although superconductivity can be induced in these regions by the
Bosonic pairs via the proximity effect, increasing doping will lead to a 
\emph{reduction }in the volume occupied by pairs and hence a rapid decrease
in $T_{c}$. The drop of $T_{c}$ in the overdoped phase may then be described
by the same PCP mechanism assuming the appearance of metallic stripes
results in a reduced volume fraction of pairs and eventual disappearance of
superconductivity at point D in Fig. 2.

In spite of the relative simplicity of the PCP model, with very few
underlying assumptions we are able to reproduce many features of the cuprate
phase diagram irrespective of the symmetry of the pairing channel. The model
gives a theoretical prediction for the ''pseudogap'' ratio $E_{p}/T_{c}$ in
terms of experimentally measurable quantities. It enables us to \emph{%
quantitatively }predict the critical carrier concentration for the
occurrence of superconductivity on the underdoped side as well as estimate
the $T_{c}$ from experimentally measured values of the pairing energy (or
pseudogap) $E_{p}$.

One of us (VVK) acknoweledge the financial support from EPSRC (UK, Grant
R46977).

Figure 1. a) the amplitude of the lattice deformation caused by pairs
described by the mesoscopic Jahn-Teller model. The picture corresponds to a
''snapshot'' at 6\% doping at $T=0$. b) the bond-percolation model
describing the situation in a).

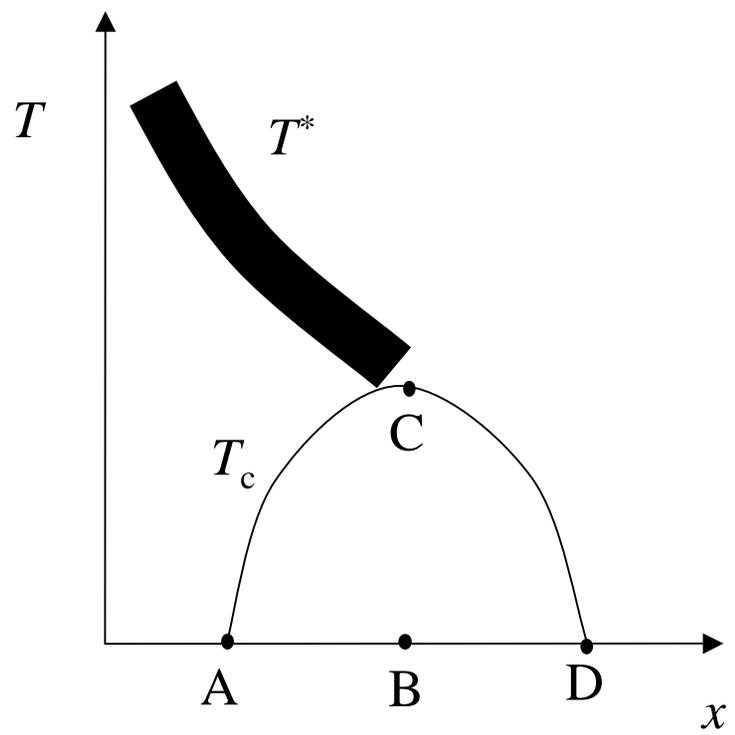
Figure 2. A schematic phase diagram of the cuprates with points referred to
in the text. The dashed line represents the pseudogap temperature $T^{\ast }$%
.

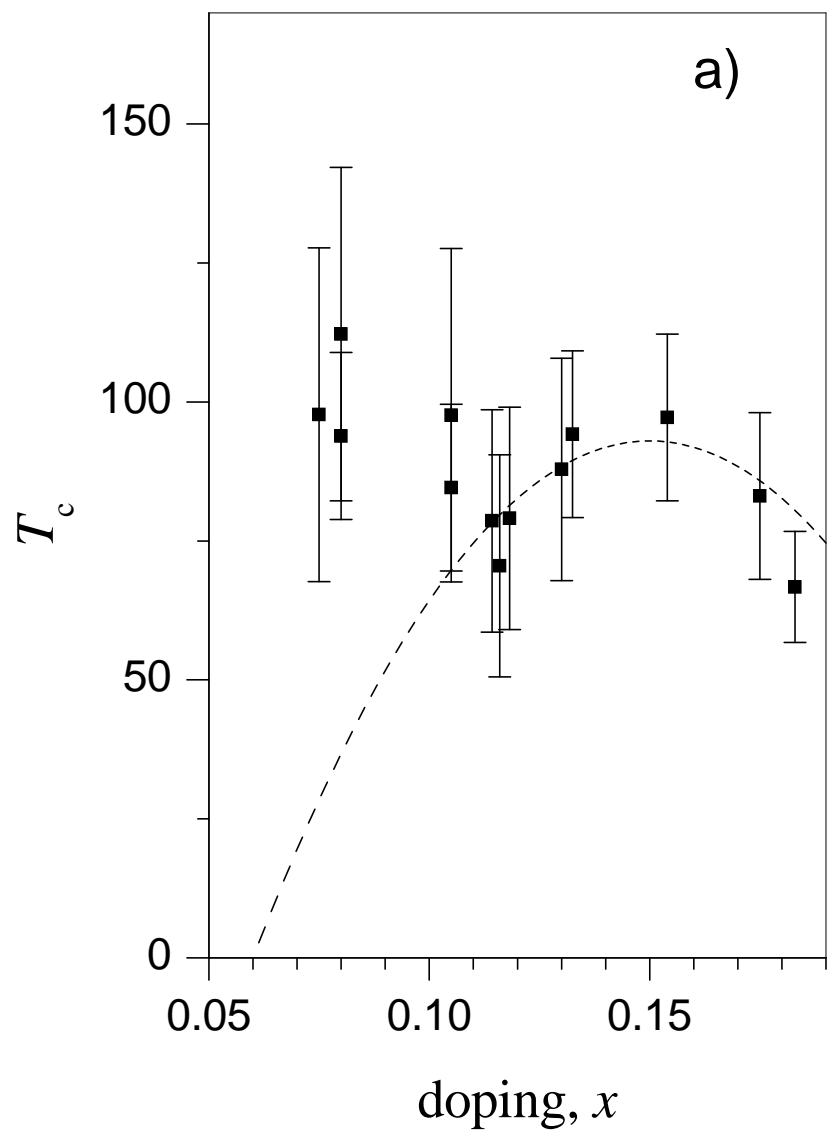
Figure 3. a) $T_{c}$ as a function of $x$ obtained from measured values of $%
E_{p}$ in YBCO. The dashed line approximately represents the measured $T_{c}$%
. b) A plot of the \emph{pseudogap ratio} $E_{p}/kT_{c}$ predicted by eeq.
(4).


\begin{thebibliography}{99}
\bibitem{tetragonal}  Here we have assumed that the crystal structure is
more or less tetragonal.
\bibitem{BEC}  If the pairs were mobile on the time-scale of $\hbar
/kT^{\ast }$, then they would form a phase-coherent state when their thermal
wavelength associated with their centre-of-mass motion is approximately
equal to the inter-particle spacing. In this case percolation would be
irrelevant and\ Bose-Einstein condensation would take place.
\end{thebibliography}
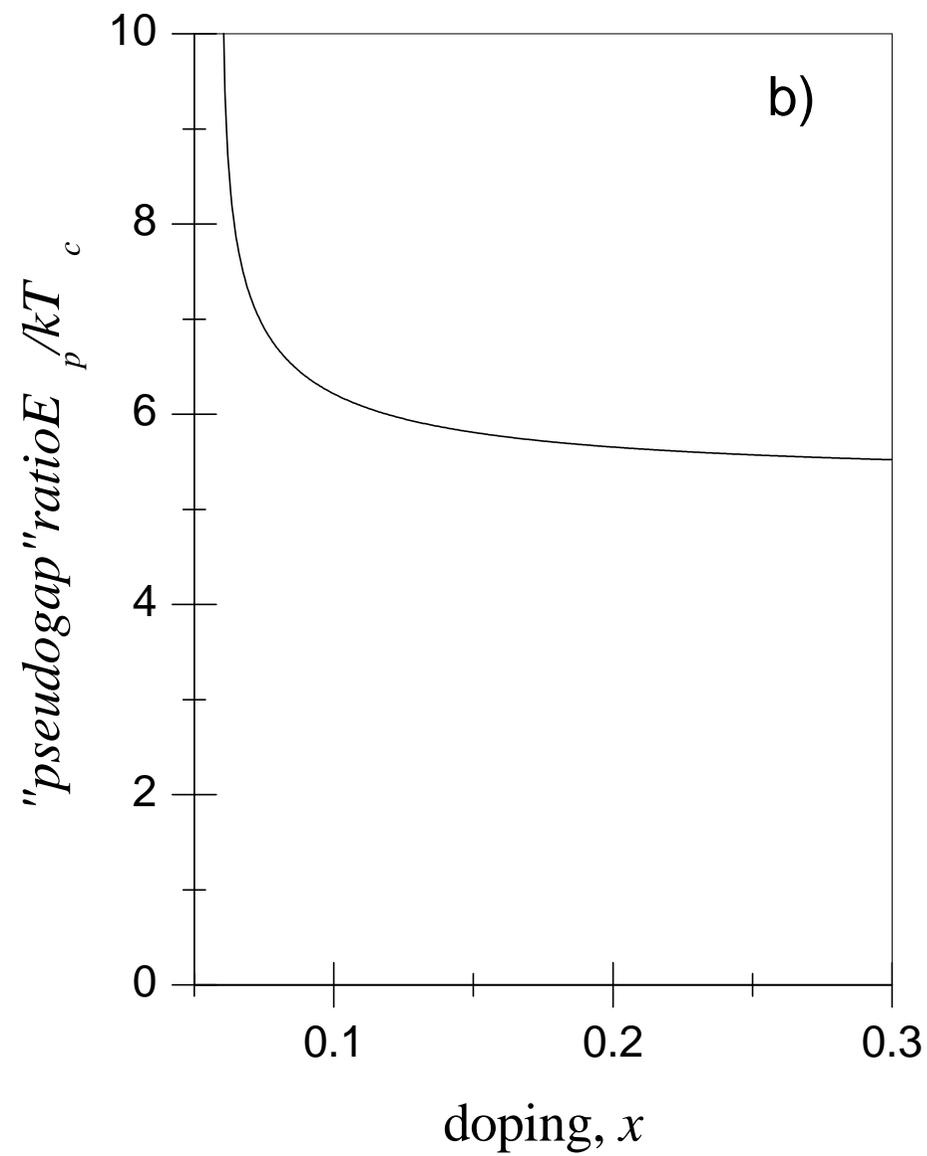
\end{document}